\documentclass[journal]{IEEEtran}
\usepackage{times}
\usepackage{cite}
\usepackage{mdwmath}
\usepackage{epsfig}
\usepackage{easymat}
\usepackage{amsfonts}
\usepackage{bm}
\usepackage{amsbsy}
\usepackage{amsmath}
\usepackage{float}
\usepackage{cite}
\usepackage{graphicx}
\usepackage{balance}
\usepackage{subfigure}
\usepackage{latexsym}
\usepackage{amssymb}
\usepackage{txfonts}
\usepackage{wasysym}
\usepackage{mathbbol}
\usepackage{multirow}
\usepackage{stfloats}
\usepackage{algorithm}
\usepackage{algpseudocode}

\usepackage{times}
\usepackage{cite}
\usepackage{mdwmath}
\usepackage{epsfig}
\usepackage{easymat}
\usepackage{amsfonts}
\usepackage{bm}
\usepackage{amsbsy}
\usepackage{amsmath}
\usepackage{float}
\usepackage{cite}
\usepackage{graphicx}
\usepackage{balance}
\usepackage{subfigure}
\usepackage{latexsym}
\usepackage{amssymb}
\usepackage{txfonts}
\usepackage{wasysym}
\usepackage{mathbbol}
\usepackage{multirow}
\usepackage{stfloats}

\begin{document}

\title{Optimized UAV Communication Protocol Based on Prior Locations}

\author{
\IEEEauthorblockN{\large Lokman Sboui  and Abdullatif Rabah}\\
\IEEEauthorblockA{
\small Electrical Engineering Program \\
\small Physical Science and Engineering (PSE) Division \\
\small King Abdullah University of Science and Technology (KAUST)
\\
\small Thuwal, Mekkah Province, Saudi Arabia \\
\small \{lokman.sboui, abdullatif.rabah\}@kaust.edu.sa
}}

\markboth{Mechatronics And Intelligent Systems,~ME 222B, May~2012}%
{L. Sboui and A. Rabah: UAV Communication Module}

\maketitle

\begin{abstract}
In this paper, we adopt a new communication protocol between the UAV and fixed on-ground nodes. This protocol tends to reduce communication power consumption by stopping communication if the channel is not good to communicate (i.e. far nodes, obstacles, etc.)\\
The communication is  performed using the XBee 868M standard and Libelium wapsmotes. Our designed protocol is based on a new communication model  that we propose in this paper. The protocole decides wether to communicate or not after computing the channel reliability through prior RSSI measurement and nodes location data

\end{abstract}


\IEEEpeerreviewmaketitle

\section{Introduction}
 \IEEEPARstart{T}{he}  unmanned aerial vehicle (UAV) is considered as an efficient tool in order to perform a specific task in various applications. However, UAV's are presenting, in general, a shortage in term of period of operation due to limited power capacities. In our application, based mainly on communication between the UAV and on ground nodes, it may be necessary that the UAV make optimal decisions either to communicate or not. The decision is based on the environment situation such as the distance between the UAV and the nearest node, presence of obstacles, etc.
The communication is performed using  an XBee 868Mhz standard and Libelium wapsmote boards. The final purpose is to perform a two-way communication between a UAV and a  Base Station (BS) through multiple nodes. The communication UAV-BS is achieved in order to acquire date related to position, speed, altitude, etc. While, the BS-UAV is used to updated data in the UAV i.e: trajectory, altitude, position, etc.\\
This paper provides an optimized protocol for communication that involves the locations of the UAV as well as the nodes. This protocol decides in each time slot whether the UAV should establish a communication or not in order to save power that may be wasted in a classical "pinging " protocol. In order to apply our protocol, a prior steps should be performed such as measuring the indoor and outdoor ranges of the antenna, defining the locations of the nodes and the path of the UAV.\\
This paper is organized as follows. In Section II, our work preliminaries are presented. Next, the
our proposed communication model is described in Section III. Communication conditions are studied in Section IV. A practical case of study is introduced in Section V.
Finally, the conclusion of the paper is presented in Section VI.

\section{Preliminaries}
%\section{Preliminaries}
In this section, used hardware is introduced then measuring RSSI is described. Finally, we present the prior data collection step.
\subsection{Communication scheme}
\begin{figure}
 \begin{center}
\includegraphics[width=4cm]{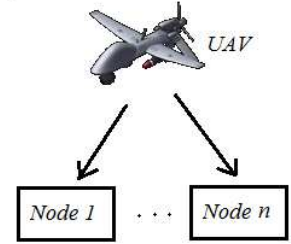}
\caption{ \label{sch} Communication scheme}
 \end{center}
\end{figure}

Our communication scheme is composed of the UAV and different nodes Figure~\ref{sch}. At each time slot the UAV have to decide either to communicate or not and with which node. Our proposed model tends to give the optimized decision in order to save power.
\subsection{Used Hardware}
\begin{figure}
 \begin{center}
\includegraphics[width=9cm]{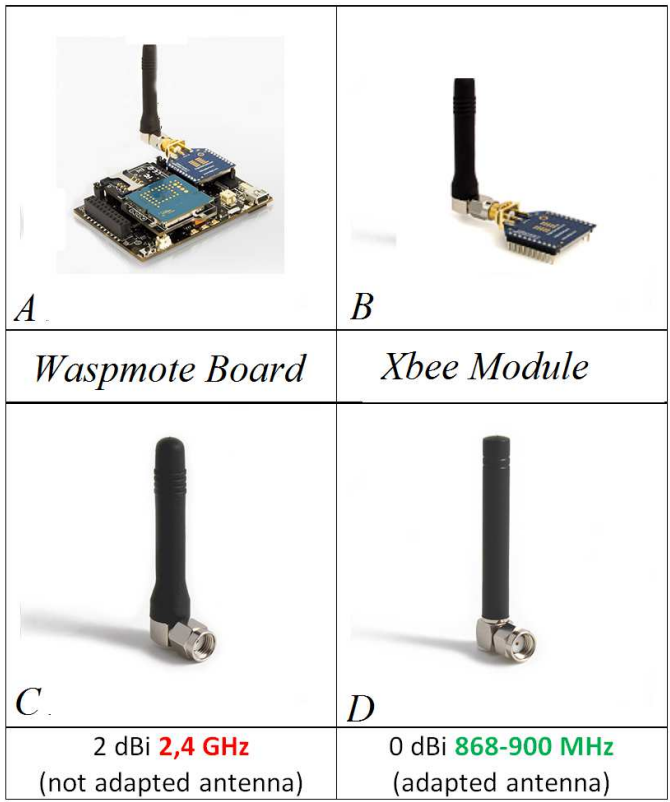}
\caption{ \label{ann} Flow chart of adopted Communication Model}
 \end{center}
\end{figure}

In our work, we use the Libelium wapsmotes~\cite{2011a}, a smart portable boards that could be used as  communication terminals Figure~\ref{ann}-A. Each waspmote is an integrated, multipurpose and microcontroller-equipped module that could be configurated in order to perform a specific task (sensing, communication, etc.)\\
The configuration is performed via a customized C code compiled and uploaded from a computer.\\
Our objective is to ensure the communication between the UAV (UAV waspmote) and the other nodes (node-wapsmotes) through the communication module (Figure~\ref{ann}-B) offered by Libelium~\cite{2012}. Available communication modules are XBee 2.4GHz, XBee 868Mhz and Bluetooth. We choose the XBee 868Mhz due to its long range up to 12Km (in theory). However, we have to make our own measurement to obtain the true ranges.
We acquired new type of antennas adapted to the 868MHz frequency band in order to obtain enhanced ranges instead of the used antennas (adapted to the 2.4GHz band).
Moreover, used antennas should be adapted to the XBee868 standard as shown in Figure~\ref{ann}-C and~\ref{ann}-D.
\subsection{RSSI Measurement}
The RSSI (Received signal strength indication) specifies, in dBm, the power strength of the received signal. The RSSI is an important measure that allows to:
\begin{itemize}
  \item Know the strength of the received power, hence the channel condition and the ability to continue the communication with the other mote.\
  \item Estimate the distance between the two motes and hence locating the mote geographically (over triangulation for example).
\end{itemize}
\subsubsection{Indoor Measurement}
We have made a number of indoor RSSI measurements and we tried to find its variation within noisy values.
Considering the fit we can associate any RSSI to a specific distance .
\subsubsection{Outdoor Measurement}
We perform a number of 4 set of measurements in 2 outdoor environments. Then, we average the obtained observation values.
We notice that for an acceptable RSSI the maximum distance is around
$700-750$ meters. \\ \\

Hence having these two curves (outdoor and indoor), we can estimate the distance from the received RSSI. For example, we know that if the RSSI is above a threshold $RSSI_0$, then we can communicate otherwise the channel is bad and we will waste energy. In this case,  if we know the distance we can decide if the RSSI is below or above the threshold then, we decide either to communicate or not.

\subsection{Data Collection}
In order to decide presence of obstacles, we have to some priori information of locations of obstacles in the region around ( UAV trajectory nearby ).
GPS coordinates of obstacles ( modeled as triangles  ) were taken from KAUST maps website : ( http://maps.kaust.edu.sa ), getting coordinates of down-left and up-right corners of each ( see figure \ref{tab} ).
We need to do transformation of GPS coordinates in order to use it in our code, because programming software doesn't supply numbers
with high significant figures. i.e. coordinate like (39.0984793763235) will be consider as ~39.0985,  which is different position:\\
Transformation is as follows (32 and 22 because our region lays on this latitude and longitude) : \\
$$ x_{new}=(x-32)\times 100000 $$
$$ y_{new}=(y-22)\times 100000 $$

\section{Communication Model}
%\section{Communication Model}
In this model, we present an algorithm that will be implemented in the mote and will optimize the UAV communication energy. The objective is to perform a communication only when the channel is in a good condition. The algorithm is a loop that runs continuously and try to figure out the nearest node to communicate and if the communication is reliable or not and then sending or receiving data.
 The  chart in Figure~\ref{chart} sums up the main steps of establishing a communication. These steps are explained as follows:
\begin{figure}
 \begin{center}
\includegraphics[width=8cm]{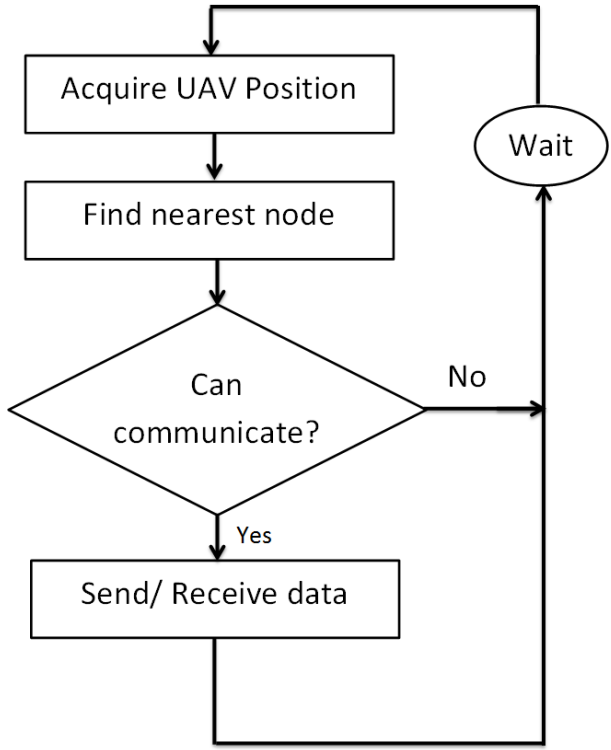}
\caption{ \label{chart} Flow chart of adopted Communication Model}
 \end{center}
\end{figure}
\subsection{Acquire UAV Position}
The position of the UAV will be acquired from the embedded GPS in the navigation system, we propose to get this position coordinates each 5 seconds in order to launch the next communication loop. In each loop, the GPS system will provide us with the Latitude, Longitude and Altitude noted $x_0,y_0 and z_0 $ respectively.
\paragraph{	Using a priori location data:}
An alternative
The principle of this method is divided into two steps:
\begin{enumerate}
  \item Follow the planned path of the UAV by car and record the GPS position for each time slot using the smartphone GPS and save these locations in a table.
  \item Include the locations table in the data base of the wapsmote and follow the path again using, in this step, the wapsmote in the same speed that we performed in step 1. Hence, we will have a pseudo real time location and the wapsmote will decide either to communicate or not.
\end{enumerate}
Note that
\begin{itemize}
  \item the GPS module could occasionally stop to operate or give an inaccurate location. If this occurs for long period we can substitute the GPS location by a triangulation method using three nodes and their corresponding RSSI such as in~\cite{Qu2010a}  where triangulation is performed using other UAV's.
  \item the GPS module was not available at the time of the experiments so we applied an alternative method to have the location.
\end{itemize}

\subsection{Find nearest node}
\begin{figure}
 \begin{center}
\includegraphics[width=8cm]{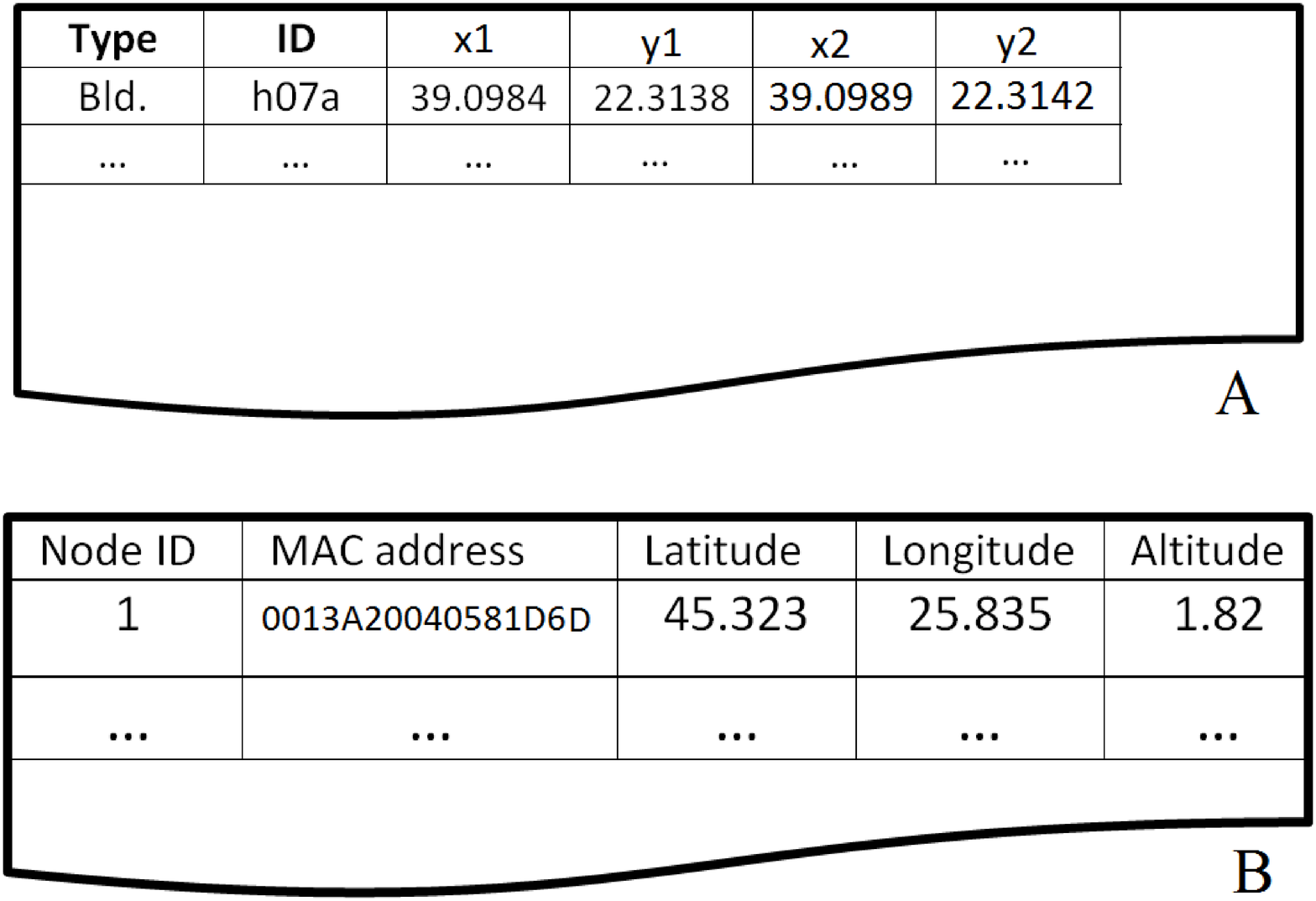}
\caption{ \label{tab} Nodes and Obstacles data bases needed in our Model}
 \end{center}
\end{figure}

We assume that the communication nodes are fixed on the ground, otherwise we have to adopt a more developed algorithm such as in~\cite{Al-Helal2010}. In addition, we assume that we have a table containing the data base of all the nodes, their positions and their MAC address (Figure~\ref{tab}-A). In each loop, the algorithm determines the nearest node by comparing the distances of the UAV position with each node and take the minimum between them.
\subsection{Communication possibility}
Once we find the nearest node we have to take the decision whether to communicate or not. First we check either the distance is good for a communication (i.e the RSSI is above the $RSSI_0$ threshold). Then, we have to make sure that the channel state is good for a reliable communication.
For this, we need a second data base of the geographical distribution of local obstacles(Figure~\ref{tab}-B). Depending on each type of obstacle and its position regarding to the direct line of sight we will take the decision either to communicate or not. Recall that adopting this algorithm will preserve more energy and maintain the UAV
operative as long as possible.
In our data base, each obstacle is defined by 5 parameters $[x, y, dx, dy, dz ]$ where$ (x,y)$ are the coordinate of the down-left corner and $[dx, dy, dz ]$are dimensions of the rectangle (width, length and height). Any point inside the obstacle will be: $( x+a\times dx ,y+b\times dy,z+c\times dz )$ where ($a$), ($b$) and ($c$) are positive number $\leqq 1$ .

\subsubsection{Identification of a direct line of sight}
Let the transmitter at point $(x_1,y_1,z_1)$ and receiver at $(x_2,y_2,z_2)$, then the equation of line between them is:
\begin{equation}\label{eq}
    \frac{x-x_1}{x_2-x_1} =   \frac{y-y_1}{y_2-y_1} =   \frac{z-z_1}{z_2-z_1}
\end{equation}
If there any obstacle inside the big rectangle defined by four points $(x_1,y_1)$, $(x_1,y_2)$,$(x_2,y_1)$ and $(x_2,y_2)$   satisfy the equation above with ($a$), ($b$) and ($c$) positive number $\leqq 1$ , then this obstacle lays in the direction of sight, hence the UAV have to decide not to establish communication.
\subsection{Send / Receive data}
Once the communication is insured, the UAV send an acknowledgment (ACK) to the nearest node in order to start the communication either by sending data (position, speed, altitude) or by receiving (updated flight path).

\section{Communication Conditions }
\begin{figure}
 \begin{center}
\includegraphics[width=9cm]{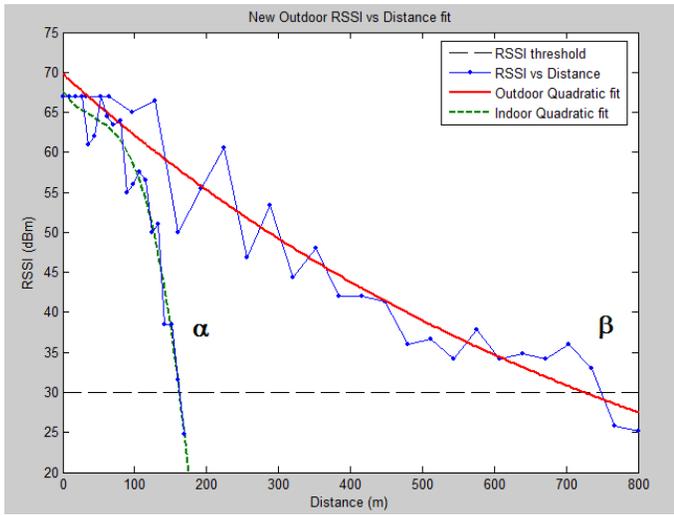}
\caption{ \label{fig_RSSI}RSSI threshold deduced by Outdoor and Indoor measurements}
 \end{center}
\end{figure}

\begin{figure}
 \begin{center}
\includegraphics[width=9cm]{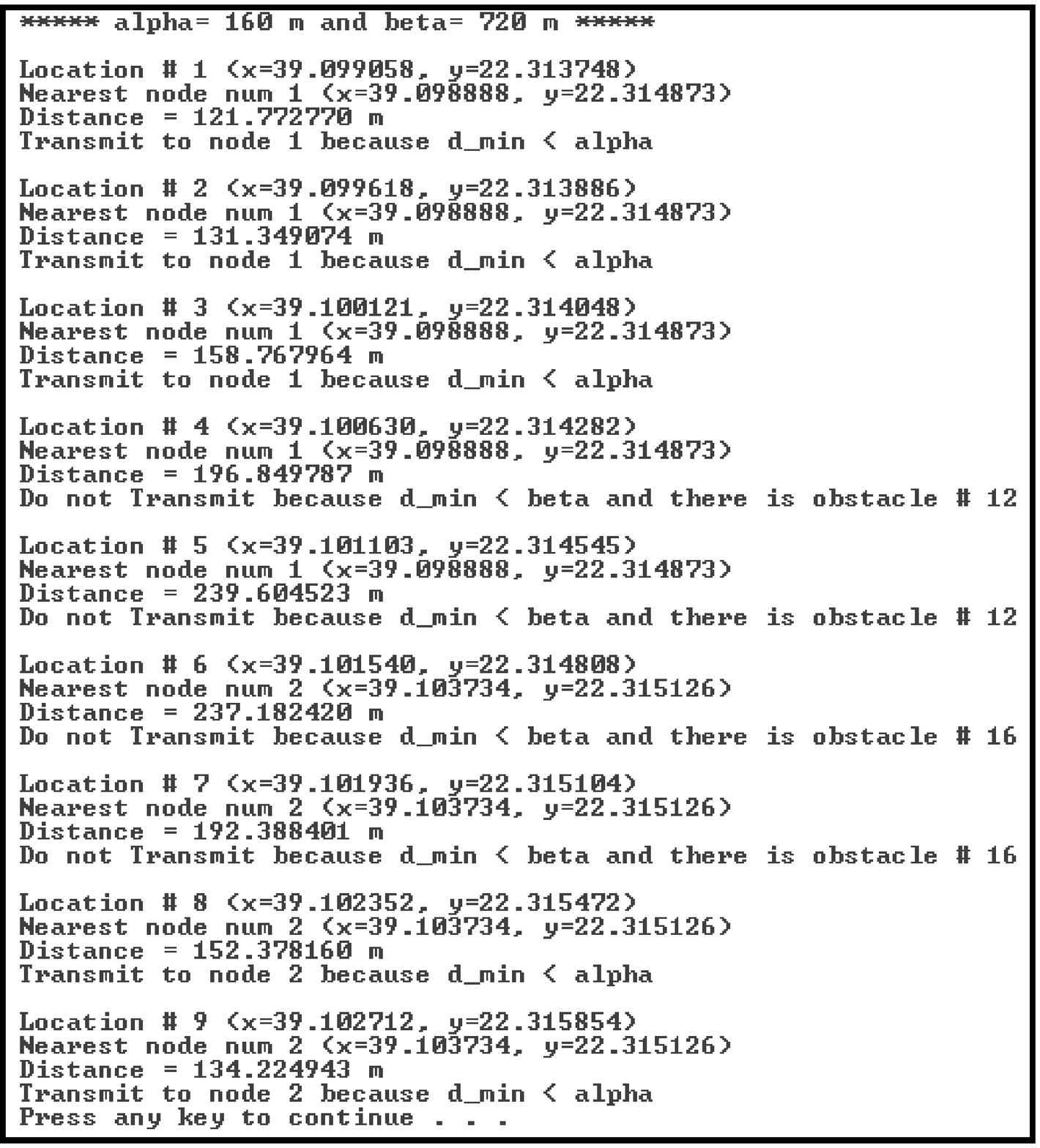}
\caption{ \label{sim}Expected communication generated by our simulation algorithm}
 \end{center}
\end{figure}

%\section{Communication Conditions}
We define some distance thresholds that determine communication possibility based on the distance between the UAV and the nearest node.
We know that the RSSI is an indicator of the channel reliability. Hence we define an RSSI threshold below which communication could not be established.
In order to insure that our communication conditions are favorable, the  RSSI threshold should also overcome the errors of GPS and RSSI measures. Otherwise we need to build an advanced filtering process such as in ~\cite{Nemra2010}. In our case, after many trials, we fix the RSSI threshold    to 30dBm.
We plot the line of (RSSI=30dBm) along with the curves of (Indoor RSSI vs. fit) and (Outdoor RSSI vs. fit), see Figure~\ref{fig_RSSI}.
We define the distance thresholds $\alpha$ and $\beta$, as the intersection between the RSSI threshold line and the indoor and outdoor curves, respectively. In this case we obtain : $\alpha=162m$ and $\beta=725m.$\\
Now, if $d_{min}$ is the distance between the UAV and the nearest node, the decision is taken as follows:
\begin{itemize}
  \item If $d_{min}<\alpha$ then "Transmit"
  \item If $d_{min}>\beta$ then "Do not Transmit"
  \item If $\alpha<d_{min}<\beta$ then "Transmit" if there are no obstacles.
\end{itemize}
We built a simulation code that acquire the positions from a pre-created table containing all the position of the UAV trajectory. For a specific case of study, we simulate the decision of the UAV using our algorithm and we obtain the results in Figure~\ref{sim}. These results match the expected decision computed manually.
This simulator is, now, a tool to predict the UAV decision before practical trails in order to make sure the communication is not stopped for long time.

 \subsection{protocol}
\begin{figure}
 \begin{center}
\includegraphics[width=9cm]{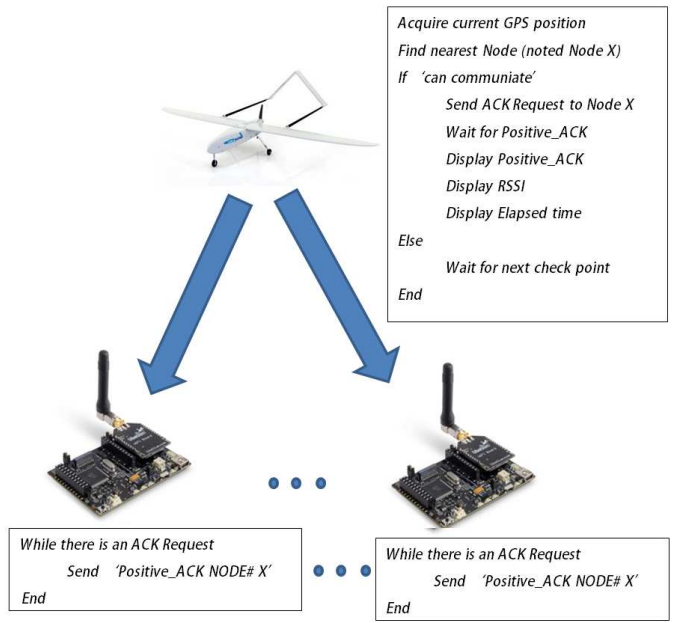}
\caption{ \label{UAV_MOTES} The proposed Communication Protocol}
 \end{center}
\end{figure}

 The protocol we propose (and adopted as you can see later, section\ref{sec4}, case of study) is shown in figure~\ref{UAV_MOTES}. Where the UAV mote is requesting acknowledgment(ACK) from nearest node. Received ACK is store (or displayed as in figure \ref{res2}) as well as RSSI and elapsed time. This is to make sure and test the availability of communication.

\section{Practical Case of Study}
%\section{Practical Case of study}
We perform a "real environment" test in order to validate the proposed communication model presented in Section III, in this test we use prior GPS locations in order to feed the UAV with the current location.\\
\subsection{Used Hardware}
 The test is performed using 4 waspmotes:
\begin{itemize}
  \item 1 mobile waspmote (UAV waspmote)  that will be attached to the UAV, in our test we use a car to simulate the mobility of the UAV.
  \item 3 fixed nodes (node-waspmotes) in order to acquire data from the UAV.
\end{itemize}
\subsection{Map}
\begin{figure*}[!t]
\includegraphics[width=18cm]{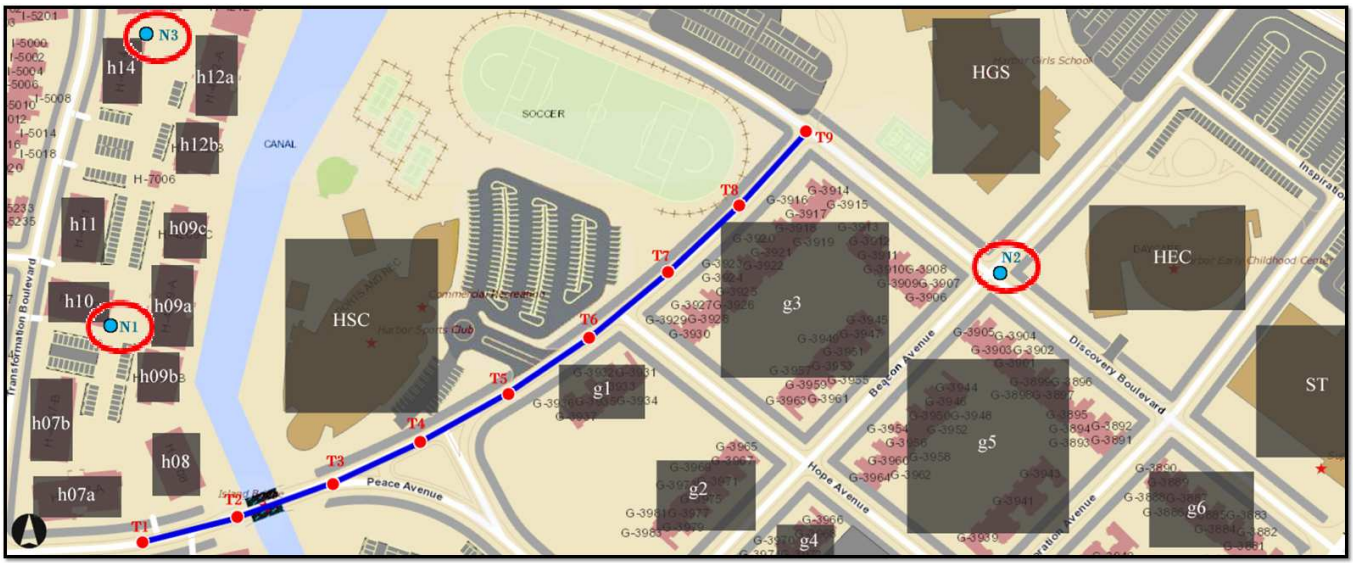}
\caption{ \label{map} The test location map at KAUST}
\hrulefill
\end{figure*}

We consider the scenario presented in the map in Figure~\ref{map}.

\subsection{Performing}
We adopt a new path (presented by the blue line) in which we feed the UAV mote by the GPS location each 50m : points (T1,...,T9). The  3 fixed nodes (N1,N2,N3)  are set in different location in order to have multiple cases of communication ( direct line of sight communication, communication within obstacle or not communicating ).
Before performing the test, we run our algorithm presented in report 4 in order to have the expected decisions. We include the new coordinates of all nodes and obstacles:
Hence, for this scenario we expect the results given in figure~\ref{sim} generated by our simulation algorithm.

\begin{figure}
 \begin{center}
\includegraphics[width=9cm]{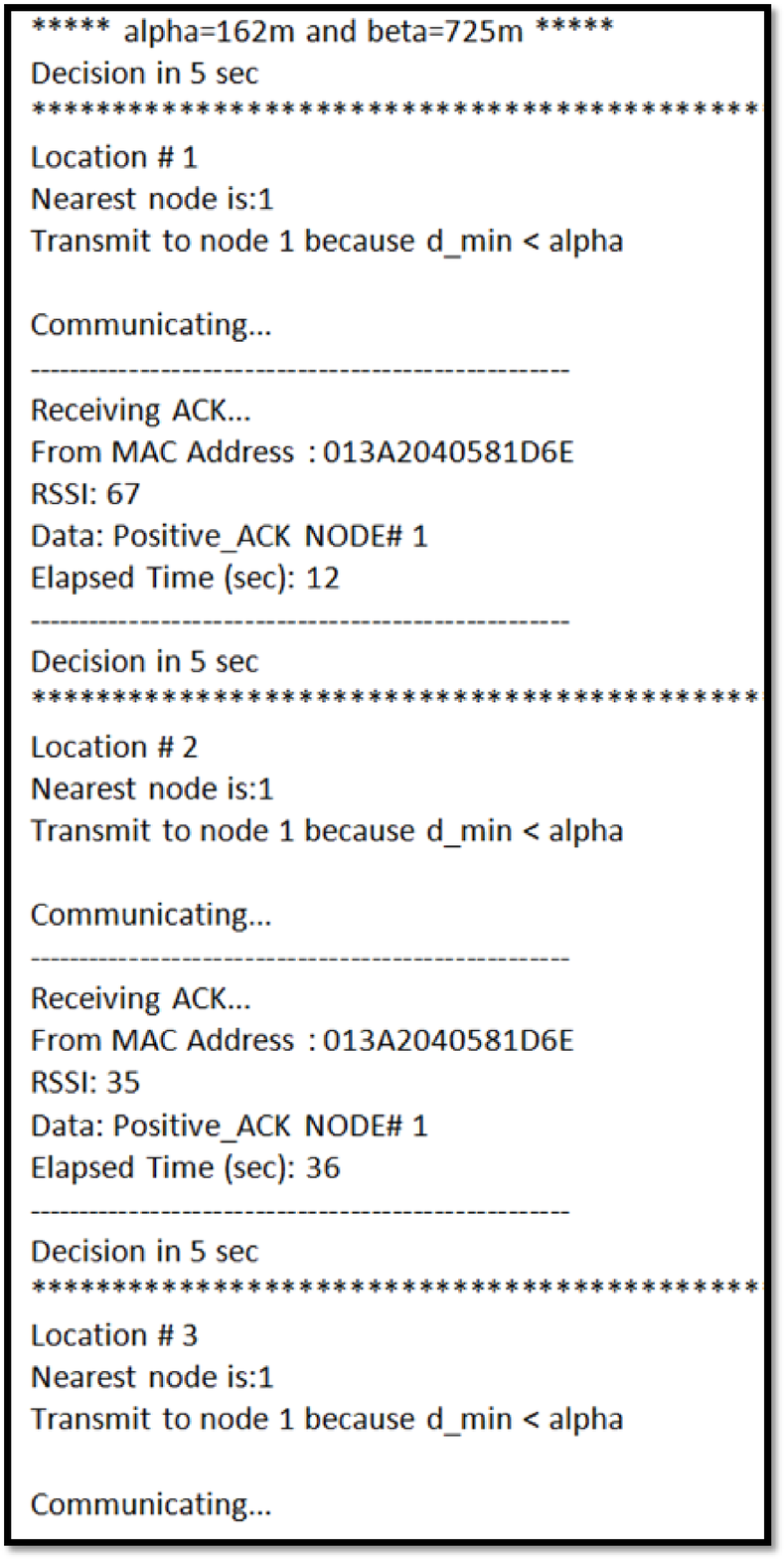}
\caption{ \label{res2} The output of the UAV wapsmote after the practical test}
 \end{center}
\end{figure}

\subsection{Results}
When running the test using a car to simulate the UAV path, we have to be in the exact step TI of the path when the decision is taken, that is why we introduced a delay of 5 sec in order to have almost exact results. We connect the UAV mote to a computer and we display its output in a monitor see Figure~\ref{res2}.

\section{Conclusion}
In this paper, we presented a new optimal communication protocol between a UAV and fixed nodes in order to save energy. This protocol is based on a communication model that uses available data ( RSSI measurement, nodes and obstacles locations) in order to give a decision wether to communicate or not. We applied our protocol in a real environment and we obtained good results, hence, our model and protocol are validated and ready to be used in the future.

%\appendices
%\section{Proof of the First   Equation}
%Appendix one text goes here.

\newpage
\section*{Acknowledgment}
The authors would like to thank Dr. Wail Gueaieb and Dr.~Christian Claudel for their advices, suggestions and discussions during this work.

\bibliographystyle{ieeetr}
\bibliography{references}

\begin{IEEEbiographynophoto}{Lokman Sboui}
received the Engineer degree from the
Department of Signals and Systems, Polytechnic Tunisia School,
Tunisia, in 2011. Currently,
he is working towards the M.S. degree at the
Department of Electrical Engineering, King Abdullah University of Science and Technology (KAUST).
His current research interests include Wireless Communications, Cognitive Radio. His publications include a conference and a journal papers in channel capacity and power allocation for Cognitive Radio systems.
\end{IEEEbiographynophoto}

\begin{IEEEbiographynophoto}{AbdulLatif Rabah}
is a master student at King Abdullah University of Science and Technology KAUST. He received his B.S. (2011) from King Fahd University of Petroleum and Minerals,Dhahran, Saudi Arabia. His intrest is in digital signal processing and its application in communications.
\end{IEEEbiographynophoto}

\end{document}